\documentstyle[aps,epsf]{revtex} 

\begin{document} 
  
\centerline{\large\bf Center vortex model for the infrared sector of}
\vspace{0.2cm}
\centerline{\large\bf Yang-Mills theory -- Topological Susceptibility}

\bigskip
\centerline{M.~Engelhardt\footnote{Supported by DFG 
under grant En 415/1--1.} }
\vspace{0.2 true cm} 
\centerline{\em Institut f\"ur Theoretische Physik, Universit\"at 
T\"ubingen }
\centerline{\em D--72076 T\"ubingen, Germany}
  
\begin{abstract}
A definition of the Pontryagin index for $SU(2)$ center vortex
world-surfaces composed of plaquettes on a hypercubic lattice is
constructed. It is used to evaluate the topological susceptibility
in a previously defined random surface model for vortices, the
parameters of which have been fixed such as to reproduce the
confinement properties of $SU(2)$ Yang-Mills theory. A prediction
for the topological susceptibility is obtained which is compatible 
with measurements of this quantity in lattice Yang-Mills theory.
\end{abstract}


\medskip

\section{Introduction}
Diverse nonperturbative phenomena characterize the infrared regime
of QCD. The basic degrees of freedom in terms of which the theory is 
formulated, quarks and gluons, are confined into color singlet hadrons.
Chiral symmetry is realized in the Goldstone mode, which decisively
influences low-energy hadronic physics. The $U_A (1)$ part of the
flavor symmetry on the other hand is affected by an anomaly \cite{thooan}
which is tied to the topological properties of the Yang-Mills fields, and
manifests itself e.g. in the mass of the corresponding flavor singlet
pseudoscalar $\eta^{\prime } $ meson \cite{witvenest}.

An in principle exact theoretical tool allowing to investigate all
these phenomena is available, namely lattice gauge theory. Nevertheless,
it is important to concomitantly develop simplified models of the
infrared sector, in order to facilitate exploratory forays into problem
areas which have proven difficult to access by conventional lattice theory
e.g. due to the numerical effort implied. Simplification can be achieved
by restricting the set of degrees of freedom and concentrating on selected
collective variables. The choice of variables should be guided on the
one hand by the question whether one can expect the corresponding effective 
dynamics to be well represented by a simple, e.g. weakly interacting, 
ansatz; on the other hand, it should be possible to tie the collective 
degrees of freedom to the nonperturbative phenomena highlighted above 
in a straightforward fashion.

An important example of such an approach is the instanton liquid model
\cite{inst},\cite{instliq}. Instantons are chosen as collective degrees
of freedom, which by construction introduces the topological properties
of the Yang-Mills fields into the model description. By virtue of the
Atiyah-Singer index theorem, the corresponding topological winding number
implies the existence of zero modes of the Dirac operator; this zero
mode spectrum is perturbed by instanton interactions into a band of 
near-zero modes, which by the Casher-Banks formula \cite{cashb} is related 
to a chiral condensate. In this manner, instanton models are very successful 
at also describing the spontaneous breaking of chiral symmetry and the 
associated low-energy hadronic phenomena. On the other hand, instanton
models have hitherto failed to provide for confinement \cite{browsh}. 
Presumably, a strongly interacting dynamics with intricate instanton 
correlations would be necessary to achieve the latter.

This example highlights that, in general, there is a conflict between
the different aims to be taken into account in formulating a model for 
infrared QCD dynamics. Also in models which have been proposed to
explain confinement, e.g. the dual superconductor \cite{tho81},\cite{mag}, 
the connection to topological properties conversely has remained tenuous;
at most, correlations have been empirically detected \cite{monumi} between, 
on the one hand, the Abelian monopole degrees of freedom of the dual 
superconductor, and, on the other hand, instantons. 

A possible exception which recently has shown promise of transcending
these difficulties is the so-called center vortex picture, which 
originally was also designed to explain confinement \cite{thoo}-\cite{spag}. 
Recent progress in the understanding of this picture 
\cite{deb97}-\cite{selprep} has, among other developments, yielded insight 
into the manner in which topological properties are encoded in the vortex 
degrees of freedom \cite{cont},\cite{cornneu}. Also first evidence has 
been gathered that center vortices indeed subsume gauge field topology 
\cite{forc1}; namely, an ensemble of Yang-Mills configurations, modified 
such as to remove all vortices from each configuration, was found to be 
concentrated on the trivial topological sector. 

In a companion paper to the present work \cite{selprep}, a simple 
model dynamics for $SU(2)$ center vortices was introduced, based on a 
representation of vortex world-sheets as random surfaces. The reader who
wishes to consult a more detailed introduction to the center vortex picture
in general, and to the physics of the aforementioned model in particular,
is referred to this companion paper. The model was treated in \cite{selprep}
on a hypercubic lattice, the random surfaces being composed of 
elementary plaquettes on the lattice, and it was shown to reproduce the 
confinement phenomenology of Yang-Mills theory. While it is not trivial 
that a simple random surface dynamics is able to qualitatively reproduce 
the confinement features (including the finite-temperature deconfinement 
transition), it should be noted that the quantitative predictive power of 
the model was not tested very stringently in \cite{selprep}. The confinement
properties essentially served to fix the parameters of the model; merely
a prediction of the spatial string tension in the deconfined phase was
possible, which turned out to be surprisingly accurate.

In the present paper, this same model, fitted to the confinement 
properties, will be used to evaluate the topological susceptibility. On 
the one hand, this represents a more solid test of the predictive 
power of the model. On the other hand, from a physical point of view, the
result will provide a further piece of quantitative evidence supporting the
conjecture that a picture based on center vortex collective degrees 
of freedom indeed is able to comprehensively and consistently capture 
the entire spectrum of nonperturbative effects characterizing the 
infrared regime of Yang-Mills theory.

\section{Pontryagin index of lattice vortex configurations}
\label{defchap}
\subsection{Continuum definition vs. lattice configurations}
In order to make the aforementioned test possible, it is first necessary to
give a workable definition of the Pontryagin index for $SU(2)$ vortex 
world-surfaces composed of plaquettes on a hypercubic lattice. This 
construction can be based on the result for general center vortices in the 
continuum, discussed in detail in \cite{cont}, which can be summarized 
as follows: The Pontryagin index $Q$ of an $SU(2)$ center vortex 
world-surface configuration $S$ is given, up to a multiplicative constant, 
in terms of its oriented self-intersection number,
\begin{equation}
Q = -\frac{1}{16} \epsilon_{\mu \nu \alpha \beta }
\int_{S} d^2 \sigma_{\alpha \beta }
\int_{S} d^2 \sigma^{\prime }_{\mu \nu }
\delta^{4} (\bar{x} (\sigma ) - \bar{x} (\sigma^{\prime } ) )
\label{yidres}
\end{equation}
where $\bar{x} (\sigma )$ denotes a parametrization of the
two-dimensional surface $S$ in four space-time dimensions, i.e. for any
two-vector $\sigma $ from a two-dimensional parameter space, the 
four-vector $\bar{x} $ gives the corresponding point on the vortex
surface in space-time. This parametrization furthermore implies an
infinitesimal surface element
\begin{equation}
d^2 \sigma_{\mu \nu } = \epsilon_{ab}
\frac{\partial \bar{x}_{\mu  } }{\partial \sigma_{a} }
\frac{\partial \bar{x}_{\nu  } }{\partial \sigma_{b} }
d^2 \sigma
\end{equation}
on the vortex surface. The normalization of (\ref{yidres})
is such that a self-intersection point in the usual sense generates a
contribution $\pm 1/2$ to the $SU(2)$ Pontryagin index, where the sign
depends on the relative orientation of the two intersecting surface
segments \cite{cont}. It should be noted, however, that (\ref{yidres}) 
not only includes contributions from intersection points in the usual sense, 
but from all singular points of a surface configuration; singular points 
in this more general sense are all points at which the tangent vectors to 
the surface span the entire four space-time dimensions. 

This result can be understood more clearly by recalling how it comes 
about physically. A vortex surface is associated with a field strength
$F_{\mu \nu } $ localized on the vortex, where the space-time directions 
labeled by $\mu $ and $\nu $ are the directions which are locally 
orthogonal to the surface. Therefore, a nonzero topological density 
$\epsilon_{\mu \nu \kappa \lambda } F_{\mu \nu } F_{\kappa \lambda } $
is generated precisely at singular points in the above general sense.
When the surface is represented by a set of plaquettes on a lattice, 
such points can occur at any site of the lattice. Note that for 
infinitely thin vortices, i.e. true two-dimensional surfaces, the field 
strength $F_{\mu \nu } $ has a singular space-time dependence.
Nevertheless, the integral over the topological density is regular
\cite{cont}. One can e.g. check this by considering an intersection
between two thickened vortices with a regular transverse profile and
a corresponding regular field strength. The resulting topological
charge is independent of the profile (as it should be for a topological
quantity) and the idealized limit of an infinitely thin vortex surface 
is therefore smooth and finite.

Note also that, via the sign of $F_{\mu \nu } $, the orientation of the
surface enters the Pontryagin index; this is a crucial new feature
compared with the evaluation of Wilson loops \cite{selprep}. Wilson
loops are insensitive to the orientation of vortex surfaces; by contrast,
to determine the Pontryagin index, one needs to give not only the location 
of the vortex surfaces, but also their orientation. This is manifest in
(\ref{yidres}) in view of the appearance of the oriented self-intersection
number. Furthermore, globally oriented surface configurations have 
vanishing self-intersection number, and thus are associated with 
vanishing Pontryagin index; non-vanishing Pontryagin index requires 
that the surfaces consist of patches of alternating 
orientation\footnote{The center vortex world-surfaces extracted from
Yang-Mills theory via center projection have indeed been found to be
non-orientable \cite{bertle} in the confined phase.}. The edges of such 
patches can be associated with monopole loops in Abelian projections 
of the gauge field representing the vortex configuration \cite{cont}.

In view of this, the following discussion of the lattice implementation 
of (\ref{yidres}) will have to be based on more specific lattice vortex 
configurations than the treatment of confinement properties \cite{selprep}; 
namely, it will be necessary to additionally introduce the notion of an 
orientation for the lattice plaquettes making up the lattice vortex 
surfaces. A suitable convention can be specified as follows. An 
orientation is associated with a plaquette by specifying a sense of curl, 
i.e. by specifying one of two possible cyclical orderings of the links 
bounding the plaquette. Formally, this can be implemented by associating 
a lattice plaquette extending from the lattice site given by the 
four-vector $n$ into the positive $\mu $ and $\nu $ directions with a 
value\footnote{Note that the superscripts $\{ \mu ,\nu \} $, where 
always $\mu \neq \nu $, are unordered sets, i.e. there is no distinction 
between $\{ \mu ,\nu \} $ and $\{ \nu ,\mu \} $.}
$p_{n}^{\{\mu ,\nu \} } $, just as in the companion paper \cite{selprep},
with a slight modification: If the plaquette is not part of a vortex surface,
$p_{n}^{\{\mu ,\nu \} } =0$. Otherwise, $p_{n}^{\{\mu ,\nu \} } =1$ if
the curl on the plaquette is given by the ordering of links 
$l^{\mu }_{n} \rightarrow l^{\nu }_{n+e_{\mu } }
\rightarrow l^{\mu }_{n+e_{\nu } } \rightarrow l^{\nu }_{n} $, whereas
$p_{n}^{\{\mu ,\nu \} } =-1$ for the opposite curl. As an illustrative
example, the following represents an oriented vortex in the shape of a
three-dimensional elementary cube surface extending into the $1,2$ and 
$3$ directions:
\begin{eqnarray}
p^{\{ 1,2 \} }_{n} = 1 \ \ \ \ \ \ & \ \ \ \ \ \
p^{\{ 1,3 \} }_{n} = -1 \ \ \ \ \ \ & \ \ \ \ \ \
p^{\{ 2,3 \} }_{n} = 1 \label{elmcub} \\
p^{\{ 1,2 \} }_{n+e_3 } = -1 \ \ \ \ \ \ & \ \ \ \ \ \
p^{\{ 1,3 \} }_{n+e_2 } = 1 \ \ \ \ \ \ & \ \ \ \ \ \
p^{\{ 2,3 \} }_{n+e_1 } = -1
\nonumber
\end{eqnarray}
Whether a link shared by two vortex plaquettes carries a monopole line
can be decided by considering in which direction that link is pursued
in the two link orderings associated with the two plaquette orientations.
If the two directions are the same, there is a monopole line; if the
two direction are opposed, the associated magnetic flux cancels.

Before proceeding to discuss how the Pontryagin index of lattice
surfaces can be defined, a note is in order concerning how ensembles
of such surfaces, containing all the necessary information, can be
generated. If the vortex surfaces are extracted from lattice Yang-Mills 
configurations, e.g. via the maximal center gauge 
\cite{deb97},\cite{deb97aug},\cite{giedt} then in view of the above 
discussion, one needs to additionally determine the locations of Abelian 
monopole lines on the vortex surfaces\footnote{The Abelian monopoles of 
the maximal Abelian gauge have empirically been determined to lie on the 
vortex surfaces of a subsequent (indirect) maximal center gauge 
\cite{deb97aug}; alternatively, in the Laplacian center gauge introduced 
in \cite{forc2},\cite{forc3}, Abelian monopoles are located on vortex 
surfaces by construction.}. Then, by associating an arbitrary orientation 
with some initial vortex plaquette on the dual lattice (this choice does 
not influence the Pontryagin index), one can recursively determine 
the orientations of all other (connected) vortex plaquettes by taking 
into account that vortex surface orientation is inverted precisely 
at monopole lines. Alternatively, if the vortex configurations are 
supplied by a random surface model \cite{selprep}, one additionally needs 
to generate orientations for the plaquettes making up the random surfaces; 
the monopole density implied by orientation changes on the surfaces can 
be viewed as an additional free parameter of such a model. This issue
will be discussed in detail in section \ref{mondiss}.

Given such oriented vortex surface configurations on a space-time lattice,
the question of evaluating their self-intersection number may at first
sight seem straightforward. In practice, however, one reencounters many of
the problems which also plague the determination of the Pontryagin index
from full lattice Yang-Mills configurations. Surfaces made up of plaquettes
on a lattice merely provide coarse-grained representations of continuum
surfaces. Accordingly, some of their features contain ambiguities as to
their precise continuum interpretation. For one, lattice surfaces in
general intersect along whole intersection lines, as opposed to the
continuum, where they intersect at points (up to a set of configurations
of measure zero). Likewise, on the lattice, monopole lines hit singular
points of the surfaces with a finite probability; by contrast, in the 
continuum, they pass by singular points at a distance (again, up to a set of 
configurations of measure zero, if one assumes monopoles to be randomly
distributed on vortex surfaces, as will be done in the following). Some
more detailed remarks on the physical meaning of the monopole trajectories
will be made in section \ref{toupsec}. The sections subsequent to the present 
one discuss in detail how the aforementioned ambiguities in the lattice 
configurations can be given a meaningful and consistent resolution, thus 
yielding a workable definition of the Pontryagin index associated with 
lattice surfaces. The resulting algorithm is summarized again in section 
\ref{algsum}.

\subsection{Removing intersection lines}
\label{remil}
Surfaces made up of plaquettes on a hypercubic lattice generically
intersect not at isolated points like in the continuum, but along whole 
lines consisting of lattice links. Along such an intersection line, even 
the question which plaquettes belong to which surface is at best a nonlocal 
question and, in general, cannot be settled unambiguously. This should be 
contrasted with intersection points, which occur at lattice sites; at such 
points, there is no ambiguity in identifying disjoint surfaces.

This ambiguity results from looking at the vortex surfaces on too coarse
a scale; if one could attain the continuum limit, one would of course
expect two-dimensional surfaces in four dimensions generically to intersect 
at points. Thus, to resolve the meaning of intersection lines, it is
necessary to specify a new surface configuration on a finer lattice, from
which the original configuration results by coarse-graining. This means
that, on the finer lattice, the surfaces may deviate from the original
surfaces by distances up to one-half of the spacing of the original 
lattice. This freedom can be used to construct the new configuration
on the finer lattice in such a way that intersection lines are
eliminated.

From the physics point of view, such a procedure is consistent. Center
vortices represent infrared degrees of freedom of Yang-Mills theory,
and are associated with a certain inherent thickness; it is not
meaningful to give their location to a higher accuracy than this
thickness. In this sense, the locations of the thin vortex surfaces 
furnished e.g. by center projection in Yang-Mills theory only carry
physical meaning up to arbitrary ultraviolet fluctuations within the
thick profile of the physical vortex they represent \cite{cont}.
Therefore, slight deformations of the thin vortex surfaces should not 
change the infrared physical content of a configuration.

To implement the above idea in practice, it is useful to introduce the
notion of an elementary cube transformation. Such a transformation,
carried out on an elementary three-dimensional cube of the lattice,
means transforming
\begin{equation}
p_{n}^{\{ \mu , \nu \} } \longrightarrow
( 1 - | p_{n}^{\{ \mu , \nu \} } | ) \cdot \bar{p}_{n}^{\{ \mu , \nu \} }
\end{equation}
for all 6 plaquettes making up the surface of the three-dimensional 
lattice cube in question, where $\bar{p} $ describes an oriented
vortex in the shape of that cube surface, such as given in (\ref{elmcub}).
Note that this transformation preserves the closedness of vortex
surfaces; furthermore, this is precisely the operation carried out in 
an elementary Monte Carlo update of a vortex configuration in the random 
vortex model \cite{selprep}, except that there, in general, $\bar{p} $ does 
not need to be oriented. In the present context, the specification that 
$\bar{p} $ be oriented has to do with the fact that it will be desirable to 
preserve the monopole content of a configuration as much as possible in the 
course of a transformation.

In practice, the ambiguities in lattice surface configurations implied by
intersection lines are now resolved as follows: Transfer the original
surface configuration onto a lattice with $1/3$ of the original lattice
spacing. Identify and make a list of all links which are part of 
intersection lines; these are all those links to which more than
two vortex plaquettes are attached. Subsequently, sweep once through
the lattice, updating the configuration by applying elementary cube 
transformations to all cubes containing links from the aforementioned list.
Accept an update if it lowers the total number of intersection line links; 
it has furthermore proven to be efficient to also accept with probability 
$1/2$ transformations which leave the number of intersection line links
unchanged. Lastly, the orientation of the corresponding cube surface 
$v$ should be chosen such as to least change the number of links carrying 
monopoles. Note that the new surfaces deviate from the original ones at 
most by $1/3$ of the original lattice spacing, in accordance with the 
requirements formulated further above.

In practice, it has proven sufficient to carry out this procedure
twice, i.e. one ends up with a lattice with $1/9$ the spacing of the
original lattice. Already after the first step, residual intersection
lines are so sparse that the second step can be carried out on local
sublattices; in this way, it is easily possible to treat $12^4 $
lattices (at the initial lattice spacing) without exceeding the
memory of a workstation. Typically, this procedure increases the vortex 
density as well as the monopole density by 15\% as compared with the
original configuration, which can be taken as an indication of
the uncertainty introduced by the procedure into measurements of
topological properties. The author has furthermore investigated
alternative procedures such as immediately going to a lattice
with $1/9$ the lattice spacing and applying several sweeps of
elementary cube transformations. These alternative procedures all
distorted the vortex and monopole densities more strongly.

\subsection{Singular points}
The procedure described in the previous section furnishes surface
configurations containing the following types of singular 
points\footnote{As already mentioned further above, singular points
are all those points at which the tangent vectors to the surface
configuration span all four dimensions.} which potentially contribute to 
the self-intersection number and, consequently, to the Pontryagin index:
\begin{itemize}
\item Touching points of two surface segments
\item Intersection points
\item Other singular points (writhing points)
\end{itemize}
The latter class of singular points is distinguished from the former
two as follows: At intersection points and touching points, two distinct 
surface segments share one point, but one cannot reach one surface 
segment from the other by proceeding along plaquettes which share a link. 
Writhing points on the other hand are characterized precisely by the 
opposite; all plaquettes attached to such a point can be connected by 
proceeding along plaquettes which share a link. In this sense, there is 
only one surface segment at such a singular point, which writhes such 
that a contribution to the self-intersection number is generated.
An explicit example of a writhing point is depicted in Fig. \ref{fig1}
below.

Also, intersection points have a very specific form. They consist
of all four plaquettes attached to the point in question and extending 
into a particular pair of space-time directions, together with all four 
plaquettes attached to the point and extending into the other pair of 
space-time directions. All other configurations at which two distinct
surface segments meet at one point constitute touching points.

All of the contributions mentioned above happen at sites of the lattice, 
which in general can also be visited by monopole lines. This represents 
an additional ambiguity which needs to be resolved in a manner similar to 
the intersection lines in the previous section. As before, in the continuum
limit, monopole lines will generically not precisely hit the singular points
of the vortex surfaces mentioned above, but pass them by at some distance
which cannot be resolved as long as the lattice spacing is finite. Thus,
a probabilistic procedure analogous to the one for intersection lines is
necessary, and will be specified in detail further below for each type
of singular point.

\subsubsection{Interplay of intersection points and writhing points}
\label{plasec}
\begin{figure}[hb]
\centerline{
\epsfysize=8.5cm
\epsffile{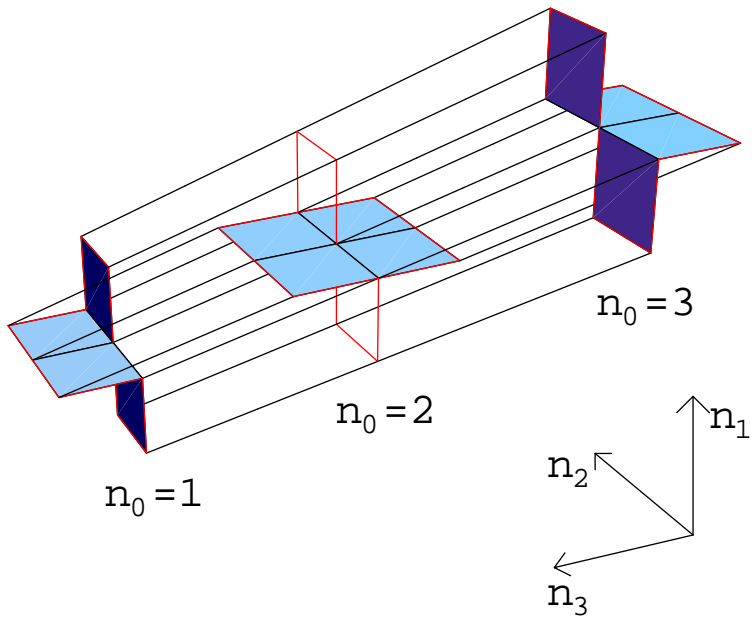}
}
\caption{Sample vortex surface configuration. At each lattice time
$n_0 $, shaded plaquettes are part of the vortex surface. These
plaquettes are furthermore connected to plaquettes running in time
direction; their location can be inferred most easily by keeping in
mind that each link of the configuration is connected to exactly
two plaquettes (i.e. the surface is closed and contains no intersection
lines). Note that the two non-shaded plaquettes at $n_0 =2$ are {\em not}
part of the vortex; only the two sets of three links bounding them are.
These are slices at $n_0 =2$ of surface segments running in time
direction from $n_0 =1$ through to $n_0 =3$. Sliced at $n_0 =2$, these
surface segments show up as lines. Furthermore, by successively assigning
orientations to all plaquettes, one can convince oneself that the 
configuration is orientable. The author gratefully acknowledges R.~Bertle 
and M.~Faber for providing the MATHEMATICA routine with which the image 
was generated.}
\label{fig1}
\end{figure}

The following discussion will repeatedly refer to the configuration
depicted in Fig. \ref{fig1}, which is helpful in illustrating the
interplay of intersection points and writhing points, and thus aids
in establishing a viable definition of the self-intersection number for
lattice configurations. 

The surface in Fig. \ref{fig1} is closed, orientable, 
has one genuine self-intersection point (precisely at the center of the 
configuration), and is devoid of intersection lines. Additionally, it has 
writhing points, at $n_0 =1$ and $n_0 =3$ in Fig. \ref{fig1}, as well as
at $n_0 =2$ at the front and back edges of the configuration (from the
viewer's perspective).

The fact that the surface is orientable implies that, when in fact oriented, 
i.e. in the absence of monopole lines, it can be represented by a gauge 
field \cite{cont} which is nonzero only on a compact region of space-time 
(namely a three-dimensional volume of which the depicted surface is the 
boundary). This implies that the existence of this configuration 
does not depend on any special boundary conditions induced by the underlying
space-time manifold and it therefore also exists e.g. on a four-dimensional
sphere. On the latter, the Pontryagin index is integer-valued. At first
sight, if one naively considered only the lone genuine intersection point to
contribute to the oriented self-intersection number, one would have
a contradiction, since each intersection point gives a contribution of
modulus $1/2$ to the Pontryagin index \cite{cont}. This underscores the 
importance of also taking into account the writhing points.

A more complete definition of the Pontryagin index can be constructed as
follows: Taking eq. (\ref{yidres}) literally, a certain unit $\gamma $ of 
topological charge should be associated with every pair of plaquettes 
which share a site and are completely orthogonal to one another, i.e. 
their combined tangent vectors span four dimensions. In terms of the 
plaquettes attached to a site $n$, the topological charge $q_n $ 
generated at that site is
\begin{equation}
q_n = \gamma \frac{1}{2}
\sum_{i,j=1}^{4} \sum_{\mu < \nu } \sum_{\kappa < \lambda }
\epsilon_{\mu \nu \kappa \lambda } \, p^{\{ \mu , \nu \} }_{(i)}
p^{\{ \kappa , \lambda \} }_{(j)}
\label{locind}
\end{equation}
where, for each pair of space-time directions, the indices $i,j$ label the 
four different plaquettes touching the site $n$ and extending into those
directions. The prefactor $1/2$ stems from the fact that the sums in
(\ref{locind}) count every pair of mutually orthogonal plaquettes twice.
Furthermore, the magnitude of $\gamma $ must be given by
\begin{equation}
\gamma =\frac{1}{32}
\end{equation}
in order to be consistent with the fact that a genuine intersection point 
of two oriented surface segments yields $q_n =\pm 16\gamma $, while it on the 
other hand is known to generate a contribution of $q_n =\pm 1/2$ to the 
Pontryagin index \cite{cont}. The sign associated with each contribution 
is determined by the relative orientation of the two plaquettes under 
consideration, as specified in (\ref{locind}). The total Pontryagin index 
$Q$ is the sum
\begin{equation}
Q = \sum_{n} q_n \ .
\end{equation}
However, one must exercise some care in applying this prescription.
For one, it is not clear what it means for a monopole line to precisely
coincide with a singular point; as mentioned above, in the continuum,
monopole lines will generically not precisely hit the singular points
of the vortex surfaces, but pass them by at some distance which cannot 
be resolved at finite lattice spacing. Therefore, monopole lines on the
lattice which happen to coincide with singular points must be deformed 
such as to circumscribe the point before applying eq. (\ref{locind}). 
There is no information contained in the configurations which would tell 
which way to deform the monopole lines, and one is therefore forced again 
to resort to a random deformation, similar to the case of the 
intersection lines in section \ref{remil}. 

In practice, this can be realized locally at, and independently for,
each site. Make a copy of the $24$ lattice plaquettes attached to
the site in question and carry out the following algorithm on this
copy: Find an initial plaquette which is part of a vortex surface, and 
proceed to find a further such plaquette sharing a link with the previous 
one. Invert its orientation if the aforementioned shared link carries
a monopole line. Iterate this procedure until the initial plaquette
is reached again. Note that up to two independent surface segments can
appear, i.e. surface segments which share nothing but the lattice
site under consideration; then, the procedure described above must
be carried out independently for both surfaces. The result of these
plaquette reorientations is that no monopole lines reach the singular 
point under consideration; to this reoriented set of plaquettes one
can now apply eq. (\ref{locind}) and thus arrive at the contribution
of the site in question to the Pontryagin index.

Note that the choice of monopole line deformation above is essentially
made when finding the initial plaquette at random. Adopting its
orientation without change means deforming the monopole line such that
it does not run across that initial plaquette. There is actually less 
ambiguity associated with this reorientation procedure than would first 
appear. In the case of writhing points, the resulting contribution to the 
Pontryagin index does not depend on the choice of monopole line 
deformation. This is due to the fact that a monopole loop encircling a 
writhing point implies inverting the orientations of all plaquettes 
attached to the point, since they are all connected via links. In other 
words, there are no two distinct surface segments which can have 
independent orientations, but there is only one surface segment. As a 
result, all pairs of orthogonal plaquettes retain their relative 
orientation, i.e. in eq. (\ref{locind}) both factors $p$ change their
sign, and the Pontryagin index is unchanged. In the case of intersection 
points, on the other hand, the choice of deformation does matter, and in 
practice is determined at random by the reorientation procedure described 
above.

The discussion in the previous paragraphs also clarifies why
eq. (\ref{locind}) could not have been naively applied already to the 
initial vortex configurations before the intersection line removal
procedure of section \ref{remil}. In the presence of intersection lines, 
the above procedure of reorienting plaquettes at a singular point such as 
to remove monopole links becomes ambiguous and nonlocal. The ambiguity 
results from the fact that intersection lines imply that there are links 
to which more than two plaquettes are attached. Therefore, when proceeding 
from plaquette to plaquette in the course of the reorientation procedure, 
there is an ambiguous choice with which plaquette to continue whenever an 
intersection line is reached. Naively, one may now think of simply making 
a random choice as in the case of other ambiguities before; however, one 
is in general not free to choose in this case. This is due to the fact 
that such a choice implies specifying which plaquettes belong to which 
surface at the intersection line. However, one then must stick to this 
choice all the way along the intersection line. At best, therefore, the 
procedure becomes nonlocal and thus unmanageable in practice. At worst, 
one may find at the opposite end of the intersection line that the choice 
was inconsistent, because there are certain classes of intersection line 
ends which do in fact imply an unambiguous partitioning of the plaquettes 
into separate surfaces.

This discussion of the conditions under which eq. (\ref{locind}) can lead 
to a sensible definition of the Pontryagin index is corroborated by the 
result one now obtains for the winding number of the configuration depicted 
in Fig. \ref{fig1}. If globally oriented, i.e. in the absence of monopole 
lines, one should obtain vanishing Pontryagin index \cite{cont}. One can
verify that this is indeed the case. Furthermore, if one additionally 
allows monopole loops on the surface, one obtains integer Pontryagin 
index. This is straightforward to understand. As explained above, monopole 
loops encircling writhing points do not change the contribution to the 
Pontryagin index generated by the point in question, since the orientations 
of all participating plaquettes are simultaneously inverted. On the other 
hand, monopole loops in general will change the sign of the contribution 
from the intersection point, e.g. a monopole loop located on one of the 
intersecting surface segments and encircling the intersection point at a 
small distance. Since the contribution from the intersection point has 
modulus $1/2$, the net change in the Pontryagin index is integer. This 
integer result is nontrivial in view of the fact that the magnitudes of 
the contributions from the writhing points in this configuration are as 
small as $1/8$. One can verify that the correct quantization of 
topological charge does not persist if one allows monopole lines to touch
the singular points. Also, by considering sample configurations
containing intersection lines, one finds that the quantization properties
of the topological charge are not preserved if one naively applies
eq. (\ref{locind}), even in the absence of monopole lines. This again
underscores the need for the probabilistic resolution of the intersection
line and monopole line ambiguities described further above, before a
consistent definition of the lattice surface Pontryagin index in terms
of eq. (\ref{locind}) can be arrived at.

\subsubsection{Touching points}
\label{toupsec}
The procedure described in the previous section accounts for the
contributions from intersection points and writhing points on a
unified footing. Some more attention is necessary with regard to
touching points. If one naively treats touching points in the same way,
they always give a vanishing contribution to the Pontryagin index
(as long as monopole lines have been deformed away from the touching
points). However, it is precisely in the latter caveat that a potential
ambiguity lies, which calls for a more detailed discussion. The
following considerations will lead to the conclusion that, even under
a less naive point of view, touching points should indeed always be 
discarded in the calculation of the lattice surface Pontryagin index.
For completeness, however, this point of view is presented here.

In the continuum, the weight of truly touching
configurations is of measure zero; instead, the following three 
continuum cases will be described by a lattice touching point:
The surfaces may intersect at two neighboring points, the surfaces
may not intersect at all, or there is only one surface, exhibiting a
thin bottleneck. Again, as long as the lattice spacing is finite, the 
configurations are too coarse-grained to allow to determine which 
possibility is realized; this represents an ambiguity in how 
surface touchings on a lattice are to be interpreted. As long as no 
monopole line is present at the same lattice site, the interpretation 
of the touching point is immaterial: The contribution to the Pontryagin 
index always vanishes, even in the case of a double intersection, since 
the two intersection points occur with opposite relative orientation of 
the surfaces \cite{cont}. The situation changes, however, if simultaneously 
a monopole line traverses the same region. {\em If} the touching point is 
interpreted as a double intersection (the interpretation would again have 
to be probabilistic) {\em and if} the monopole line is interpreted as 
running between the two intersection points (again with a certain 
probability), then the two contributions of modulus $1/2$ to the 
Pontryagin index from the two intersection points do not cancel, but add, 
changing the index by an integer \cite{cont}.

In order to discuss whether this is a meaningful interpretation, it is
necessary to return to the underlying physical picture \cite{selprep}.
The vortex surface collective degrees of freedom treated here are meant
to represent the infrared information contained in thick physical center
vortices, as already mentioned further above. To specify
that such a vortex carries flux corresponding to a center element of the
gauge group is a statement about the magnitude of the associated gauge
field color vector, but not about its direction in color space. As one
travels along a physical thick vortex, the associated gauge field
will in general smoothly rotate in color space. In concentrating on the
vortex surfaces treated here, consisting of patches of alternating
orientation, one has idealized the vortices in two ways: For one, the
flux of the physical vortex has been concentrated on thin two-dimensional
surfaces. Furthermore, however, the initially smooth space-time dependence 
of the associated color vector has been replaced by discrete jumps, allowing
the color vector to only point into the positive and negative Cartan
direction(s) in color space, as encoded in the surface orientation.
This can be interpreted in terms of casting the initial physical vortex
into an Abelian gauge and performing Abelian projection.

In view of this physical picture, it becomes clear that the interpretation
presented above, generating additional contributions to the Pontryagin
index from surface touching points, overstretches the applicability of
the idealized vortex surface degrees of freedom discussed here.
If a vortex surface touching point is interpreted as a double intersection,
then the two intersection points are very close together compared with
the typical length scales of the theory on which color vectors 
vary\footnote{A quantitative measure of this length scale is furnished
by the mean monopole density.}. Therefore, the color orientations of the
participating physical vortex surfaces should never be considered to
vary appreciably between the two intersection points; this means that
monopole lines should always be deformed such as to circumscribe
touching points. Otherwise, one accords their specific space-time
location more physical significance than is appropriate for infrared
effective degrees of freedom. The monopole lines are merely intended
to encode the variation of vortex color vectors on infrared length
scales. 

As a result of this physical interpretation, touching points should indeed be 
simply discarded in the calculation of the lattice surface Pontryagin index.

\subsection{Summary of the algorithm}
\label{algsum}
The definition of the lattice surface Pontryagin index constructed above
can be briefly summarized as follows:
\begin{itemize}
\item Transfer the given lattice surface configuration onto a lattice
of $1/3$ the lattice spacing and sweep once through the lattice, applying
elementary cube transformations such as to remove intersection lines,
while conserving the number of monopole lines as much as possible.
Carry out this whole procedure twice.
\item For each lattice site $n$, make a copy of all attached plaquettes,
and transform the copy as follows. After finding an initial plaquette
which is part of a vortex, recursively reorient further vortex plaquettes 
sharing links with previously considered ones, such as to remove any
monopole lines from the shared links. Do this for all independent surface 
segments present. Using the transformed plaquettes, obtain the contribution 
$q_n $ to the Pontryagin index from the site $n$ in question via 
eq. (\ref{locind}), i.e.
\begin{displaymath}
q_n = \frac{1}{64}
\sum_{i,j=1}^{4} \sum_{\mu < \nu } \sum_{\kappa < \lambda }
\epsilon_{\mu \nu \kappa \lambda } \, p^{\{ \mu , \nu \} }_{(i)}
p^{\{ \kappa , \lambda \} }_{(j)}
\end{displaymath}
\item The total Pontryagin index is the sum $Q=\sum_{n} q_n $.
\end{itemize}
Despite the formal similarity of the expression for $Q$ with the
so-called field theoretical discretization of the Pontryagin index in
lattice Yang-Mills theory \cite{digic}, a fundamental difference must
be emphasized. The latter definition approximates the field strength
by lattice Yang-Mills plaquettes and thus contains a discretization
error as long as the lattice spacing is finite; eq. (\ref{locind}) by
contrast {\em contains no such discretization error}. While the vortex
surfaces are composed of elementary squares, they can trivially be
embedded into a space-time continuum, and the above definition then 
already represents the true continuum Pontryagin index associated
with the vortex configuration. Thus, the algorithm given above is
conceptually more akin to the so-called geometrical definition of
the Pontryagin index \cite{luescher}. Also there, coarse-graining
ambiguities in the lattice configurations have to be resolved, before
then applying an in turn exact prescription of evaluating the Pontryagin
index.

Applying this procedure to general surface configurations on a hypercubic
lattice, i.e. a space-time torus, one obtains a Pontryagin index quantized
in half-integer units. This is exactly as it should be \cite{baal}, in
contrast e.g. to a space-time sphere, on which the Pontryagin index must be
an integer. To further corroborate the significance of this result, the
author has empirically considered some surface configurations with
Pontryagin index $1/2$ in detail. All of them turned out to be
non-orientable, meaning that they necessarily contain monopole
loops, and therefore nontrivial gauge fields at space-time infinity.
Conversely, if one formally removes the monopoles and introduces
double vortex surfaces (which are equivalent to Dirac string world-sheets
\cite{cont}) spanning areas bounded by the monopole loops, then the 
Pontryagin index again becomes an integer. In terms of the underlying gauge 
fields, this operation corresponds to a gauge transformation up to 
singularities at space-time infinity, the neglection of which is presumably 
equivalent to switching twist sectors on the torus \cite{baal}. 
Accordingly, it seems plausible to expect that certain classes of 
non-orientable vortex surfaces do not exist e.g. on a space-time sphere 
in the sense that one cannot construct gauge field representations for 
them which fulfill viable space-time boundary conditions. 

These observations further corroborate the consistency of the present 
construction of the lattice surface Pontryagin index. It should be noted
that the vortex ensembles extracted from Yang-Mills theory via center
projection and Abelian projection (to locate the monopoles) in general
contain all twist sectors, even if the original full Yang-Mills ensemble
was constrained to a particular sector. This can happen due to the fact that
the projection procedures truncate gauge-invariant information, including
in general the torus twist. Also in this respect, the vortex formulation
of the topological winding number has similar problems as e.g. standard
cooling techniques, in which instantons can ``fall through the lattice'',
modifying the detected topological content.

\section{Topological properties of the random vortex model of infrared
Yang-Mills theory}
\label{mondiss}
In the following, measurements of the topological susceptibility in the 
framework of the random vortex model of infrared Yang-Mills theory will be 
presented. This model was defined in \cite{selprep} on the basis of the
unoriented plaquette degrees of freedom $|p_n^{\{ \mu , \nu \} } |$ by 
the partition function
\begin{equation}
Z = \left( \prod_{n} \prod_{\mu ,\nu \atop \mu < \nu } \ \
\sum_{|p_n^{\{ \mu , \nu \} } |=0}^{1} \right)
\Delta [\, | p_n^{\{ \mu , \nu \} } | ] 
\exp (-S [\, | p_n^{\{ \mu , \nu \} } | ] )
\label{partit}
\end{equation}
with the constraint
\begin{eqnarray}
\Delta [\, | p_n^{\{ \mu , \nu \} } | ] &=&
\prod_{n} \prod_{\mu } \delta_{L_{n}^{\mu } \bmod 2 ,0} \\
L_{n}^{\mu } &=& \sum_{\nu \atop \nu \neq \mu } \left(
| p_n^{\{ \mu , \nu \} } | + | p_{n-e_{\nu } }^{\{ \mu , \nu \} } | \right)
\end{eqnarray}
enforcing closedness of the vortex surfaces by constraining the number
$L_{n}^{\mu } $ of vortex plaquettes attached to the link extending from
the lattice site $n$ in $\mu $ direction to be even, for any $n$ and $\mu $.
This constraint is conveniently enforced in Monte Carlo simulations by
allowing only updates via elementary cube transformations, 
cf. section \ref{remil}. The action penalizes curvature in the 
surfaces\footnote{In \cite{selprep}, also an action term proportional 
to the total vortex surface area was considered, but a good fit to the 
confinement properties of $SU(2)$ Yang-Mills theory was achieved by 
choosing the corresponding coupling constant to vanish.} by considering, 
for every link extending from the lattice site $n$ in $\mu $ direction, 
all pairs of attached plaquettes whose two members do not lie in the 
same plane, and, if both members are part of a vortex, associating this 
with an action increment $c$:
\begin{equation}
S = \frac{c}{2} \sum_{n} \sum_{\mu } \sum_{\nu , \lambda \atop
\nu \neq \mu , \lambda \neq \mu , \lambda \neq \nu }
\left( | p_n^{\{ \mu , \nu \} } p_n^{\{ \mu , \lambda \} } |
+ | p_n^{\{ \mu , \nu \} } p_{n-e_{\lambda } }^{\{ \mu , \lambda \} } |
+ | p_{n-e_{\nu } }^{\{ \mu , \nu \} } p_n^{\{ \mu , \lambda \} } |
+ | p_{n-e_{\nu } }^{\{ \mu , \nu \} }
p_{n-e_{\lambda } }^{\{ \mu , \lambda \} } | \right) 
\label{cdef}
\end{equation}
The confinement properties of the $SU(2)$ Yang-Mills ensemble are best
reproduced by the choice $c=0.24$, cf. \cite{selprep}. This to a large
extent fixes the characteristics of the random surface ensemble.
However, as is manifest in the above definition by the fact that the
model is formulated in terms of the unoriented variables 
$| p_n^{\{ \mu , \nu \} } |$, the orientations of the surfaces 
remain undetermined; this is simply a consequence of the fact
that Wilson loops, which encode the confinement properties, do not depend 
on them. By constrast, the topological properties can only be evaluated 
by taking the surface orientation into account. One thus needs to 
additionally generate orientations for the plaquettes describing the 
vortex configurations which make up the ensemble.

In practice, this was achieved by first assigning random orientations to
the plaquettes of any given configuration from the ensemble (\ref{partit});
subsequently, Monte Carlo sweeps were performed through the configuration
in which plaquette orientations were flipped according to a bias related
to the monopole number on the links bounding the plaquette in question.
In this way, different monopole densities can be generated on the given
vortex surfaces. Note that this density can only be varied within
certain bounds. On the one hand, the non-orientability of the surfaces 
implies a certain irreducible monopole density; on the other hand, the 
lattice spacing also provides an upper bound on the density of monopoles.

At zero temperature, the above procedure allowed to study monopole line
densities in space-time within the range\footnote{The scale in the
following is fixed by setting the zero-temperature string tension 
$\sigma_{0} $, measured in \cite{selprep}, to 
$\sigma_{0} = (440 \, \mbox{MeV} )^2 $.}
\begin{equation}
\frac{6.2 \, \mbox{fm} }{\mbox{fm}^{4} } < \rho (T=0) 
< \frac{38 \, \mbox{fm} }{\mbox{fm}^{4} } \ .
\label{monrho}
\end{equation}
For comparison, the zero-temperature monopole line density measured in full 
$SU(2)$ Yang-Mills theory in the maximal Abelian gauge \cite{borny} 
amounts to $\rho_{MAG} = 64/\mbox{fm}^{3} $. This latter value 
should certainly be regarded as too high for the purposes of the random 
surface ensemble, for the same reasons as the center projection vortex 
density in Yang-Mills theory is higher than the model vortex density, 
by about a factor two. This difference, discussed extensively in
\cite{selprep}, is due to spurious ultraviolet fluctuations in the
center projection vortices detected from the Yang-Mills ensemble.
Since monopole loops are constrained to lie on vortex surfaces, their
density likewise should be reduced by at least a similar
factor. Indeed, the magnitude of $\rho_{MAG} /2$ is within the upper 
bound in (\ref{monrho}). 

The random surface ensemble can also be studied at finite temperatures
\cite{selprep} by varying the extension of the lattice universe in the
Euclidean time direction, this extension being identified with the inverse 
temperature. The parameter choice $c=0.24$ in the action (\ref{cdef}) in 
particular \cite{selprep} induces the correct $SU(2)$ Yang-Mills relation 
between the deconfinement phase transition temperature $T_C $ and the 
zero-temperature string tension $\sigma_{0} $, namely 
$T_C /\sqrt{\sigma_{0} } =0.69$. Temperatures in the following can thus 
consistently be given in units of $T_C $. Similar to above, at high 
temperatures, the range of accessible monopole densities 
e.g. at $T=1.66 T_C $ was
\begin{equation}
2/\mbox{fm}^{3} < \rho (T=1.66 T_C ) 
< 14/\mbox{fm}^{3} \ .
\label{monrhot}
\end{equation}

\begin{figure}
\centerline{
\epsfysize=8.3cm
\epsffile{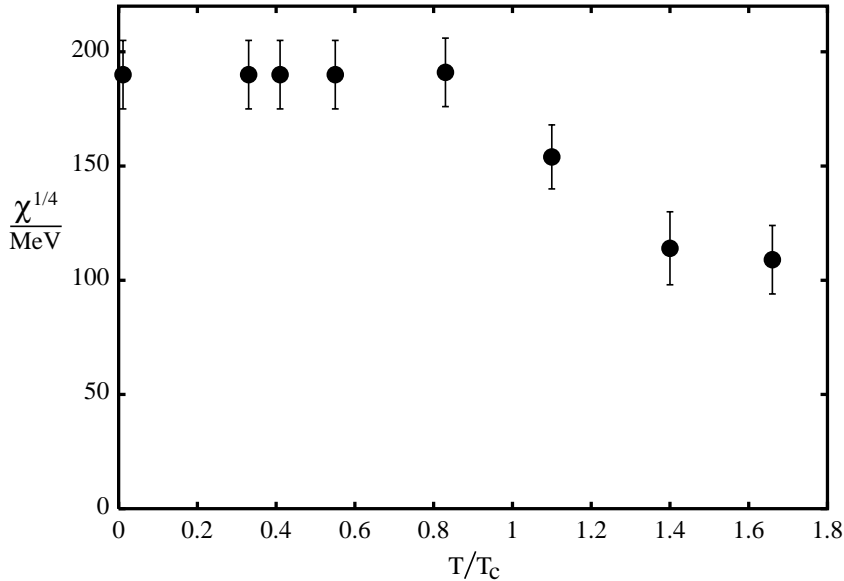}
}
\caption{Topological susceptibility as a function of temperature in
the random vortex model. The random surface ensemble (\ref{partit})
was generated on $12^3 \times N_t $ lattices and was supplemented
by (biased) generation of plaquette orientations (see text). To obtain 
the values at $T=1.1 T_C $ and $T=1.4 T_C $, an interpolation procedure 
was used. This procedure, described in detail in \protect\cite{selprep},
takes recourse to alternative values of $c$ in the action (\ref{cdef})
which allow to generate the aforementioned temperatures, whereupon
interpolation in $c$ allows to define the observable also for the
desired value $c=0.24$.}
\label{chifig}
\end{figure}

One might a priori expect that these monopole densities can be tuned such 
that the random surface model reproduces the topological susceptibility of 
Yang-Mills theory. However, when measuring the topological susceptibility
in diverse ensembles distinguished only by their monopole densities, the
variation of the result turned out to be smaller than the error bars
over the entire range of accessible densities. Thus, the random surface 
model in actual fact gives a unique prediction of the topological 
susceptibility independent of any additional model tuning. The main 
reason for this lies in the fact that writhing points are considerably 
more abundant than intersection points; however, as discussed in section 
\ref{plasec}, the contribution generated by the former can be evaluated 
independently, and is invariant under changes, of the monopole 
configuration\footnote{This does not mean to say that the topological 
charge generated by writhings {\em exists} independently of any monopole 
loops. Rather, the non-orientability implied by the writhings necessarily 
also implies a certain irreducible monopole density, but the topological 
charge associated with writhings can be evaluated without explicit knowledge 
of these monopoles.} (as far as they are topologically possible). As an
example, while the full zero-temperature topological susceptibility
amounts to $\chi^{1/4} (T=0) = (190\pm 15)$ MeV, cf. Fig. \ref{chifig},
the zero-temperature susceptibility generated by taking only
intersection points into account is
$\chi_{int}^{1/4} (T=0) = (110\pm 10)$ MeV, again over the whole range
(\ref{monrho})). This at first sight surprisingly stable result is presumably 
due to the additional effect that the degree of non-orientability of 
the vortex surfaces, which implies a lower bound on the density of 
monopoles, already suffices to effectively randomize the signs of the 
contributions $\pm 1/2$ from intersection points to the Pontryagin index; 
as a consequence, additional monopole loops do not further appreciably 
change the variance of the topological charge associated with these 
intersection points.

Similarly, at high temperatures, the topological susceptibility
generated by intersection points alone e.g. at $T=1.66 T_C $ amounts to 
$\chi_{int}^{1/4} (T=1.66 T_C) = (21\pm 8)$ MeV, over the whole range
(\ref{monrhot}). In this case, intersection points are very sparse;
more than 95\% of the configurations contain no intersection points. 
As evidenced by comparison with the full value of the topological 
susceptibility, $\chi^{1/4} (T=1.66 T_C) = (109\pm 10)$ MeV, writhing 
points also dominate in the high-temperature phase. Therefore, also in 
this phase, one obtains a unique prediction for the topological 
susceptibility.

Fig. \ref{chifig} displays the topological susceptibility $\chi $ in the
random surface model as a function of temperature. The zero-temperature
value, $\chi^{1/4} (T=0) = (190\pm 15)$ MeV, is compatible with other
measurements in full $SU(2)$ Yang-Mills theory \cite{digic}. The
topological susceptibility persists in the deconfined phase, but drops to
$\chi^{1/4} (T=1.66 T_C) = (109\pm 10)$ MeV at $T=1.66 T_C $. The
temperature dependence is similar to the behavior displayed in
\cite{digic}, but the error bars do not quite overlap at $T=1.4 T_C $
and $T=1.66 T_C $. However, as far as the author understands, the
error bars in \cite{digic} are purely statistical, and no estimate is
included of the systematic error inherent in the extraction procedure 
used. Thus, it is not clear at this stage whether the residual deviation
at high temperature signals a genuine discrepancy between the random 
vortex model and full Yang-Mills theory.

The chief uncertainty in the measurements displayed in Fig. \ref{chifig}
indeed results from the systematic error introduced by the modification 
of the vortex surfaces after they have been transferred onto a finer 
lattice, cf. section \ref{remil}. The magnitude of this error is estimated 
from the change induced by the aforementioned procedure in quantities 
such as the vortex and monopole densities. However, at high temperatures, 
statistical errors become comparable to the systematic uncertainties due 
to a stronger autocorrelation in the course of the ensemble generation.

In terms of the physical consequences, while the $U_A (1)$ anomaly
persists in the deconfined phase, it loses its dominating character
e.g. in the Witten-Veneziano estimate \cite{witvenest} for the mass of 
the $\eta^{\prime } $ meson. Rather, it becomes one effect among
several, comparable e.g. in magnitude to the explicit $SU(3)$ flavor
symmetry breaking via the quark masses.

\section{Conclusions and Outlook}
In this work, a random surface model for $SU(2)$ center vortices, designed
to reproduce the confinement properties of Yang-Mills theory \cite{selprep},
was shown to simultaneously predict the correct topological susceptibility.
While confinement and the topological aspects of the Yang-Mills ensemble
have hitherto largely been modeled on a separate footing, the vortex
picture bridges this gap and the aforementioned phenomena find a
unified description. The center vortex framework thus shows increasing
promise of being suited to consistently capture the gamut of
nonperturbative effects characterizing the infrared regime of the
strong interactions.

In order to arrive at this result, a workable definition of the Pontryagin
index for vortex world-surface configurations composed of plaquettes on a 
hypercubic lattice had to be developed, cf. section \ref{defchap}. Similar 
to the so-called geometrical definition of the Pontryagin index 
\cite{luescher} in lattice Yang-Mills theory, coarse-graining ambiguities 
in the lattice configurations must be resolved before the known continuum
expression \cite{cont} for the Pontryagin index can be used.
Work is in progress with the aim of applying these methods also to
the center projection vortices extracted from the Yang-Mills ensemble.
This will allow to explore the question of vortex dominance for the
Pontryagin index, in analogy to the center dominance observed for
Wilson loops \cite{deb97},\cite{temp},\cite{tlang}. Such an investigation
constitutes a more stringent test than the converse experiment \cite{forc1},
which showed that an ensemble devoid of center vortices is concentrated in 
the trivial topological sector.

Besides confinement and topological properties, the vortex model will
have to be confronted with the phenomenon of spontaneous chiral symmetry
breaking. For this purpose, fermionic degrees of freedom must for the
first time be included into the description. On a technical level,
it will be necessary to develop efficient methods of evaluating the 
Dirac operator in a vortex background. In this respect, the vortex model 
presents a simplification compared with full lattice QCD which may be 
decisive: While the random surfaces describing the model vortex ensemble 
(\ref{partit}) are composed of lattice plaquettes, there is no
conceptual difficulty in embedding them in a space-time continuum;
indeed, one can straightforwardly associate a continuum field strength
and a continuum gauge field with the surfaces. The advantages connected 
with this feature to some extent already became apparent in the 
construction of the Pontryagin index, cf. section \ref{defchap}, and
it also opens the possibility of a continuum evaluation of the Dirac
operator. From a physical aspect, the success of instanton models in
generating the spontaneous breaking of chiral symmetry leads to the
expectation that the vortex model will fare likewise, in view of its
correct description of the topological characteristics highlighted
in this work.

\section{Acknowledgments}
First and foremost, the author wishes to thank M.~Faber for numerous
intense and fruitful discussions during the author's stay at TU Vienna
in September 1999, where the basic ideas of the present work were
developed. The hospitality of the Institute of Nuclear Physics at 
TU Vienna as a whole is acknowledged, as is DFG for financing the
stay within grant no. DFG En 415/1-1. The author furthermore profited from
insights provided by H.~Reinhardt, in collaboration with whom prior work 
\cite{cont},\cite{selprep} leading up to the present one was carried out.
Also, useful discussions with R.~Alkofer, R.~Bertle, \v{S}.~Olejn{\'\i}k, 
M.~Quandt and O.~Schr\"oder contributed to this investigation.

\end{document}